\baselineskip=16pt
\normallineskip=6pt
\vsize=26 true cm
\hsize=16 true cm
\voffset=-1.0 true cm
\font\bigfont=cmr10 scaled\magstep1
\font\Bigfont=cmr10 scaled\magstep2
\font\ninerm=cmr9
\footline={\hss\tenrm\folio\hss}
\pageno=1
\newcount\fignumber
\fignumber=0
\def\fig#1#2{\advance\fignumber by1
 \midinsert \vskip#1truecm \hsize14truecm
 \baselineskip=15pt \noindent
 {\ninerm {\bf Figure \the\fignumber} #2}\endinsert}

\def\ref#1{$^{[#1]}$}
\def\sqr#1#2{{\vcenter{\vbox{\hrule height.#2pt
   \hbox{\vrule width.#2pt height#1pt \kern#1pt
   \vrule width.#2pt}\hrule height.#2pt}}}}

\def\lan{\langle }
\def\ran{\rangle }
\bigskip
\bigskip
\centerline{\Bigfont  Analytic Determination of the Complex-Field Zeros of REM}
\bigskip
\smallskip
\medskip
\centerline{\bigfont C.~Moukarzel and N.~Parga}
\centerline{\it Centro At\'omico Bariloche}
\centerline{\it 8400 S.C.~de Bariloche, R\'\i o Negro, Argentina}
\centerline{\it email:\bf cristian@if.uff.br}
\medskip
\bigskip
\noindent{\bf Abstract} \par
The complex field  zeros  of  the  Random  Energy  Model  are
analytically determined. For $T<T_c$  they  are  distributed  in  the
whole complex plane with a density that decays very fast with  the
real component of H. For $T>T_c$ a region is found which is  free  of
zeros and is associated to the paramagnetic phase of  the  system.
This region is separated from the former by a line of zeros.
The real axis is found to be enclosed in a cloud of zeros  in the
whole frozen phase of the system.
\par \bigskip \bigskip \noindent
PACS:  75.10.Nr (Spin Glasses), 05.20.-y (Stat.Mech.)
\par \bigskip \bigskip \bigskip
\noindent
{\bf 1. Introduction}
\par
\smallskip
          The determination of the zeros of the  partition  function
allows much insight into the behavior of a model.  Since  their
relation with phase transitions were clarified by the works of Yang
and Lee~[1,2] and M. Fisher~[3],  many  interesting  properties  of
zeros distributions have been studied in homogeneous systems~[4-9].
\par
More limited, on the other hand, are the  results  concerning
zeros distributions of disordered models. Taking into account that
all the information about a given model is contained in the  zeros
of its partition function, it is reasonable to expect that many of
the interesting properties~[10]  of  disordered  systems  will  be
somehow reflected by them.
\par
An example of this can be found in the complex magnetic field
zeros of spin glasses, which are {\it dense} near the  real  axis.  This
property has been suggested by Parisi [11] as being a  consequence
of the multiplicity of states and was recently illustrated~[17] by
a numerical calculation of the complex field zeros distribution in
the Random Energy Model (REM)~[12].
\par
Here an analytical calculation of the complex field zeros  of
REM is reported. The technique is the same as recently employed by
Derrida [13] to calculate the complex $\beta $ zeros of the  same  model,
which had been also numerically estimated~[14,15].
\par
In section 2 we describe the basics of the REM  and  give  an
account of the ideas that allow  the  calculation  of  its  zeros.
Section 3 is devoted to  the  REM  in  a  complex  field  and  the
calculation of the dominant contribution to its partition function
in each region of the complex $h =\beta H$ plane.
In section 4 these  results are used to obtain the density of zeros
of  REM  in  the  complex field plane, while section 5 contains our
conclusions.
\medskip
\noindent 2.   {\bf THE RANDOM ENERGY MODEL}
\par
\smallskip
The REM is defined~[12]  to  have $2^N$
independent random energy levels
$E_i$,  i=1,..$2^N$
distributed  according to
$P(E) = (\pi N)^{-1/2} \exp (-E^2/N)$.
The partition function of  a  {\it sample}
(a sample is a set of $2^N$ numbers $E_i$), \
${\cal Z}  =  \sum_{i=1}^{2^N} e^{-\beta E_i}$
can  be  written as
${\cal Z} = \sum_{E} n(E) \  e^{-\beta E}$
where $n(E)$ is the number of levels  between $E$ and $E+ \Delta E$.
This density of states $n(E)$ if found to satisfy
\medskip
$$
\cases{\lan n(E)\ran    \sim   \exp N (\log 2-({E \over N})^2)
  \cr  \cr
\lan n(E)^2\ran - \lan n(E)\ran ^2    \sim
  \lan n(E)\ran \cr }
\eqno(1)
$$
\medskip
\noindent
so two different energy regions can be identified:
\par
If $|E|< N \epsilon _c$ ,  with $\epsilon_c=(\log 2)^{1/2}$,
then  the  number  of states $n(E)$ is much greater than  one
and  its  fluctuations  are small, so for a typical sample
we will have~[13]
\par
$$
n^{\hbox{typ}}(E) =\lan n(E)\ran
+ \eta _E \lan n(E)\ran ^{1/2} \eqno(2)
$$
\noindent where $\eta _E$ is a random number of order 1.
\par
On the other hand, if $|E| > N \epsilon _c$ ,
then the number of  states
is exponentially small meaning that a typical sample will have  no
levels in this region
\par
$$
n^{\hbox{typ}}(E) = 0
\eqno(3)
$$
\noindent
The partition function for a typical sample will then be
\par
\medskip
$$
{\cal Z}^{\hbox{typ}} = A + B  \eqno(4)
$$
\noindent
with
\par
\medskip
$$
A = \sum_{|E| < N \epsilon _c} \lan n(E)\ran e^{-\beta E}
\eqno(5)
$$
\medskip
\noindent
and
\par
\medskip
$$
B = \sum_{|E| < N \epsilon _c} \eta_E \lan n(E)\ran^{1/2} e^{-\beta E}
\eqno(6)
$$
\medskip
\noindent
The calculation of $A$ may be done by the  steepest
descent  method~[16] for complex $\beta  =  \beta _1 + i \beta _2$.
\par \medskip
$$
A~  \sim ~2^N \int^{+\epsilon _c}_{-\epsilon _c}
 \hbox{d}e \quad \exp { - N (e^2 + \beta e)  }
\eqno(7)
$$
\noindent The integration contour is stretched to infinity through
paths  of steepest descent  for  the  real  part  of  the  exponent
of  the integrand.
When the integration limits lay on  opposite  sides  of
the saddle point in $e = -\beta /2$ we can write
\par
\medskip
$$
\int^{+\epsilon _c} _{-\epsilon _c}                  =
\int^{-\infty }     _{-\epsilon _c}                  +
\int^{+\infty }     _{-\infty }                      +
\int^{+\epsilon _c} _{+\infty }                   \eqno(8)
$$
\medskip
\noindent and $A$ has contributions both from the saddle point
and  from  the integration   limits   for    large    N.
This    happens    if
 $|\beta _1| < \beta _c = 2 \epsilon _c = 2 (\log 2)^{1/2}$
and we get
\par
\medskip
$$
A~\sim ~\exp {\lbrace N \beta \epsilon_c        \rbrace }
      + \exp {\lbrace N (\log 2 + (\beta /2)^2) \rbrace }
\eqno(9)
$$
\noindent whereas if  $|\beta _1|  > \beta _c$ the
two steepest descent paths run  from  the
integration limits down to infinity on the same side of the saddle
point so it does not contribute and we have
\par
\medskip
$$
\int^{+\epsilon _c} _{-\epsilon _c}                   =
\int^{\pm \infty }  _{-\epsilon _c}                  +
\int^{+\epsilon _c} _{\pm \infty }                   \eqno(10)
$$
\medskip
\noindent so
\par
$$
A~\sim ~\exp { \lbrace {N\beta \epsilon_c} \rbrace }
\eqno(11)
$$
\medskip
\noindent
It has to be mentioned~[13] that the lowest lying  states  in  REM
have order one fluctuations from sample  to  sample~[12]  so  the
contributions coming from the integration limits, both in (9)  and
(11), are known up to a fluctuating  complex  (if $\beta $  is  complex)
factor of order one. We will return to this point later on because
it results crucial for our calculation.
\par
The contribution $B$ of the fluctuations can be calculated  for
large N. Owing to the randomness of $\eta _{E}, B$ is not the
integral  of
an analytic function so the steepest descent method is  no  longer
applicable. On  the  other  hand,  the $\eta _{E}$  are
uncorrelated  for
different values of $E$ so the term with the  largest  modulus  will
dominate the sum in (6). We can then estimate the modulus of $B$ as
\par
\medskip
$$
|B|   \sim   \max_{-N \epsilon_c \le E \le N \epsilon_c}
| \lan n(E)\ran^{1/2} e^{-\beta E}~|
\eqno(12)
$$
\medskip
\noindent
It is easily seen that when $|\beta _1| <\beta_c/2$
the maximum is in  between the limits so
\par
$$
|B|  \sim \exp {\lbrace {N\over 2} ( \log 2 + \beta^2_1) \rbrace  }
\eqno(13)
$$
\noindent
while if $|\beta_1| > \beta_c /2 $ one of the limits dominates and
\par
$$
|B| \sim \exp {\lbrace { N|\beta_1|\epsilon_c)} \rbrace }  \eqno(14)
$$
\noindent
We will now use these results to calculate ${\cal Z}$ for
a typical  sample of REM when a complex field $H$ acts on it.
We will take $\beta $ real from now on, as well as $h_1$ and $h_2$
non-negative $(h=\beta H=h_1+ih_2)$
which can be done without any loss of generality.
\par
\medskip
\noindent
3.   {\bf REM IN A COMPLEX FIELD}
\par
\noindent
When a magnetic field $H$ acts on the system, the energies $E_i$ of the
states now have, in addition to the internal energies (coming from
interactions among spins) a contribution from the interaction with the
field, $E_i = U_i - H M_i$, where the $M_i$ are the magnetizations of
the states. The partition function ${\cal Z} =\sum_{i}\
 e^{-\beta  E_i}$
is then
\par
\medskip
$$
{\cal Z}(\beta ,h) =\sum_{i=1}^{2^N} \quad e^{-\beta  U_i  + h M_i}
\eqno(15)
$$
\medskip
\noindent
The internal energies $U_i$ have a Gaussian distribution which is
independent of the field, so taking into account that among the
$2^N$ states of  the  system, $g_m = g_M ={ N \choose {(N+M)\over 2}}$
have magnetization $M = N m$, we can write
\par
$$
{\cal Z}(\beta ,h)~=\sum_{-1 \le m \le 1 } e^{Nmh} ~ {\cal Z}_m(\beta )
\eqno(16)
$$
\noindent
where
\par
\medskip
$${\cal Z}_m(\beta )  =  \sum_{i=1}^{g_m}\quad e^{-\beta U_i}
 \eqno(17)
 $$
\medskip
\noindent
is the partition function of a sample of REM with $g_m$ instead of
$2^N$ states. If we write $g_m = ( q_m)^N$, then  all  the  results
of  last section concerning ${\cal Z}$ are valid for ${\cal Z}_{m}$
with the sole  condition  of replacing  $\log 2$  by
$$
\log q_m = \log 2 - {1\over 2} \left[(1+m)\log (1+m) +
 (1-m)\log (1-m)\right]
$$ \noindent
(for large $N$).
\par   \noindent
These restricted partition functions ${\cal Z}_m$ can be written as
\par
$$
{\cal Z}_m(\beta )  = \sum_{-\infty < u < \infty } n(u,m)~e^{-N \beta u}
\eqno(18)
$$
\noindent
where $n(u,m)$ are the numbers of states with internal energy $Nu$ and
magnetization $Nm$. They can be seen to satisfy
\par
\medskip
$$ \cases{\lan n(u,m)\ran ~\sim ~ \exp N (\log q_m - u^2) \cr
  \cr
\lan n(u,m)^2\ran - \lan n(u,m)\ran^2~\sim ~\lan n(u,m)\ran \cr }
\eqno(19)
$$
\medskip
 
\medskip
\noindent
Then for a typical sample we will have
\par
\medskip
$$
\cases{ n^{\hbox{typ}}(u,m)\sim \lan n(u,m)\ran +
\eta (u,m) \lan n(u,m)\ran^{1/2} \hfill \qquad
\hbox{,if}\quad |u| < u_c(m) \cr
 n^{\hbox{typ}}(u,m)\sim 0 \hfill \hbox{,if}\quad |u| > u_c(m) }
\eqno(20)
$$
\medskip
\noindent
where  the $\eta (u,m)$  are  random  numbers  describing the  typical
fluctuations, and the energy threshold $u_c(m)$ is now $m$-dependent.
\par
$$
u_c(m) = ( \log  q_m )^{1/2}    \eqno(21)
$$
\noindent
The partition function of a typical sample under field will then be
\par
$$
{\cal Z}^{\hbox{typ}}(\beta ,h) = \sum_m e^{Nmh}
{\cal Z}^{\hbox{typ}}_m (\beta )   \eqno(22)
$$
\noindent
We now use the results of last section and write
\par
$$
{\cal Z}^{\hbox{typ}}_m(\beta ) =   A_m (\beta ) + B_m(\beta )
\eqno(23)
$$
\noindent
with
\par
\medskip
$$
\cases {A_m = \rho_m \exp { \lbrace N \beta u_c (m) \rbrace }
\hfill \qquad
\hbox{, if}\quad \beta > 2 u_c(m)                   \cr   \cr
 A_m = \rho_m \exp { \lbrace N \beta u_c (m) \rbrace }
 + \exp { \lbrace N(\log q_m +(\beta /2)^2 ) \rbrace }
 \qquad \hfill
 \hbox{, if}\quad \beta < 2 u_c(m)  }      \eqno(24)
$$
\medskip
\noindent
and
\par
\medskip
$$
\cases { B_m = \eta_m \exp { \lbrace N \beta u_c (m)\rbrace }
\qquad \hfill \hbox{,if}\quad
 \beta > u_c(m)    \cr \cr
 B_m = \eta_m  \exp { \lbrace {N \over 2} (\log q_m + \beta^2 )\rbrace }
 \qquad \hfill
 \hbox{,if}\quad \beta <  u_c(m)    }         \eqno(25)
$$
\medskip
\noindent
Here $\eta _m$ is a random number of order one which describes the
fluctuations in the density of states as  already  mentioned.  The
origin of $\rho _m$ is somewhat different. They are due to fluctuations
in the values of the  lowest  lying  states: The ground state
energy (internal energy in this case) $U_0$ has order one sample to
sample fluctuations~[12,13] so $U_0^{\hbox{typ}}(m) \sim N (u_c(m)
 + \zeta /N )$ with $\zeta $ an order one  random  number  so
 $e^{\beta U_0} \sim  \rho e^{N\beta u_c(m)}$, where $\rho $ is a
positive fluctuating number.
\par
We see in this way that both in $A$  and $B$  the  ground  state
contributions are non-analytic (due to  disorder),  and  therefore
non-integrable by the steepest descent method. The  only  analytic
contribution is the saddle-point term in (24).
\par
Let us now turn to the evaluation of  the  sum  in  (22).  We
define $\beta_c(m)$ as
\par
$$
\beta_c(m) = 2 u_c(m)       \eqno(26)
$$
\noindent
and its inverse $m_c(\beta )$ such that satisfies
\par
$$
\beta /2 =\left[ \log 2 - {1\over 2}\left[(1+mc)\log (1+mc)
+ (1-mc)\log (1-mc)\right] \right]^{1/2}       \eqno(27)
$$
\noindent
The relation $h_c(\beta ) = \hbox{atanh}(m_c(\beta ) )$ defines the
critical field $h_c$ which is the limit between the frozen
and paramagnetic phases  of REM with field [12].
This critical field satisfies
\par
$$
\beta /2 = \left[ \log 2 + \log (\cosh ( h_c))-h_c
\tanh (h_c)\right]^{1/2}
\eqno(28)
$$
\noindent
Using (26) we can rewrite the condition $\beta >2u_c(m)$
as $|m|>m_c(\beta )$  and the condition  $\beta >u_c(m)$
as $|m|>m_c(2\beta )$. The sum in (22) can now be written
\par
\medskip
$$
{\cal Z}^{\hbox{typ}} = A  +  B      \eqno(29)
$$
\medskip
\noindent
with
\par
$$
A  = \sum_m  A_m(\beta ) e^{Nmh}  \eqno(30)
$$
\noindent
and
\par
$$
B = \sum_m B_m(\beta ) e^{Nmh}        \eqno(31)
$$
\noindent
We then have
\par
\medskip
$$
 A = A_1 + A_2 = \sum_{-1\le m\le 1} \rho_m e^{N[\beta u_c(m)+mh]}  +
 \sum_{ |m| < m_c(\beta )} e^{N[(\beta /2)^2 + \log q_m +mh]} \eqno(32)
$$
\medskip
$$
 B = B_1 + B_2 = \sum_{ |m| > m_c(2\beta )}
 \eta_m e^{N[\beta u_c(m)+mh]}  +
 \sum_{ |m| < m_c(2\beta )} \eta_m e^{{N\over 2}[\beta^2
 + \log q_m +2mh]}                                      \eqno(33)
$$
\medskip
\noindent
The only analytic contribution is $A_2$, which will be  estimated  by
means of the steepest descent method. We will ignore $B_1$ because it
is already included in $A_1$. Let us first discuss the evaluation  of
the fluctuating contributions $A_1$ and $B_2$.  The  non-analyticity of
the integrand makes it impossible  to  use  the  steepest  descent
method. On the other hand, as already discussed in section II, the
term with the greatest modulus will dominate the sums so we have
\par
\medskip
$$
| A_1| \sim  \max_{-1 < m < 1 }   \quad
 \exp { \lbrace N\lbrack \beta  (\log q_m)^{1/2}+ mh_1\rbrack \rbrace }
 \eqno(34)
$$
\medskip
\noindent
and
\par
\medskip
$$
|B_2|\sim \max_{|m| < m_c(2\beta )} \quad
 \exp {\lbrace {N\over 2}\lbrack \beta^2  (\log  q_m )
 + 2 m h_1 \rbrack \rbrace }
\eqno(35)
$$
\medskip
\noindent
It is worth noticing that
\par
$$
|B_2|=0\qquad \hbox{if }\beta  > \beta_c/2      \eqno(36)
$$
\noindent
On the other hand, if $\beta < \beta_c/2 $,then $m_c(2\beta ) > 0$
and  there  exist two ranges of $h_1$ in which $B_2$ gives different
contributions:
\par \smallskip
The maximum of $|B_2|$ is at $\bar{m} = \tanh ( 2h_1)$  when
$|\bar{m}|< m_c(2\beta )$ and at $\pm m_c(2\beta )$ when
$|\bar{m}|> m_c(2\beta )$.
This can be written, using  (27) and (28) as
\par
\medskip
$$
|B_2| \sim
\cases{ \exp {\lbrace {N\over 2}\lbrack \beta^2+\log 2
+\log (\cosh (2h_1))\rbrack \rbrace }
\qquad \hfill \hbox{,if}\quad |h_1| < {1\over 2} h_c(2\beta )  \cr \cr
\exp {\lbrace N\lbrack \beta^2 + |h_1| m_c(2\beta )\rbrack \rbrace }
\hfill  \hbox{,if}\quad |h_1| > {1\over 2} h_c(2 \beta )   } \eqno(37)
$$
\medskip
\noindent
Now we discuss the evaluation of $A_1$. The maximization condition is
satisfied at $\hat{m} = \tanh (\hat{h})$ where $\hat{h}$ is a function
of $h_1$ and $\beta $ that satisfies
\par
\medskip
$$
{\beta \over 2}{\hat h\over  h_1} = \left[ \log 2 + \log
\cosh (\hat h) -  \hat h \tanh (\hat h) \right]^{1/2}       \eqno(38)
$$
\medskip
\noindent
If we put $\hat h = h_c(\hat{\beta })$ with $\hat{\beta } = a\beta $
, now  the  unknown  is $a(\beta ,h_1)$.
Replacing this in (38) we find $\hat h= a h_1$, so the following
set of equations:
\par
\medskip
$$
\cases{ {\hat{\beta } / 2} = \lbrack \log 2 + \log \cosh (\hat h)
- \hat h \tanh (\hat h) \rbrack^{1/2} \cr
\hat\beta h=\beta \hat h              }  \eqno(39)
$$
\medskip
\noindent
is equivalent to (38).
\par
It is easy to see that when $h_1 = h_c(\beta )$ we obtain $a =1$  while
$a<1$ if $h_1>h_c(\beta )$ and $a>1$ if $h_1<h_c(\beta )$, so $\hat h$
will  always  be between $h_1$ and $h_c(\beta )$.
\par
\noindent
The term $A_1$ will then have the following form
\par
$$
|A_1|\sim \exp { \lbrace N \lbrack {\hat{\beta }\beta \over 2}
+h_1\tanh(\hat h) \rbrack \rbrace }            \eqno(40)
$$
\noindent
The calculation of $A_2$ may be done by the steepest descent method
(see the appendix for details) for large $N$. It follows that again
two different expressions are found, now depending on the values of $h$
and $\beta $. A smooth arc is found in the complex $h$ plane which
divides these two possibilities. This arc touches the real axis at
$h_c (\beta )$ and the imaginary axis at $h = i \pi /2$.
In the inner part of this arc we find
\par
\medskip
$$
A_2 \sim \exp N \{{\beta^2\over 2} + h m_c(\beta )  \} +
\exp  N \{ {\beta^2+ \beta_c^2 \over 4} + \log (\cosh (h)) \}
\eqno(41)
$$
\medskip
\noindent
and in the outer side
\par
\medskip
$$
A_2 \sim \exp N \{{\beta^2\over 2} + h m_c(\beta )  \}
\eqno(42)
$$
\medskip
\noindent
It is clear that ${\cal Z}$ will in general be the sum of
several terms  of
the form $e^{N\varphi }$ with $\varphi $ a complex function, so for large
$N$ just  that term with the greatest modulus will be relevant.
We then  have to determine, in each region of the complex $h$ plane,
the  term  with the greatest real part of $\varphi $. This will give rise
to the appearance of a phase diagram for the model in complex h.
\par
In  what  follows  we  will be only interested in the determination of
the modulus of ${\cal Z}$, which, as we shall see, is enough to calculate
the density of zeros of the model.
If we write $\mid {\cal Z}\mid =\exp N\phi $, the different contributions
we have to compare are
\par
$$
\eqalign{
\quad &\phi_1={\beta^2+ \beta_c^2 \over 4}+\log |\cosh (h)|
\qquad \qquad
\hfill
\hbox{for} \quad \beta <\beta_c\quad \hbox{and} \quad |h_1|
< h_1^{\hbox{lim}}(h_2,\beta )
\cr
\quad &\phi_2={\beta^2 \over 2}+|h_1|m_c(\beta )
\qquad  \qquad \qquad \qquad \quad
\hbox{for}\quad \beta <\beta_c\quad \hbox{and}\quad \forall h_1
\cr
\quad &\phi_3={\beta \hat{\beta }\over 2}+h_1\tanh (\hat h)
\qquad \qquad \qquad \qquad  \qquad  \quad
\hfill
\forall \beta  \quad \hbox{and}\quad \forall h_1
\cr
\quad &\phi_4={\beta^2\over 2}+{\beta_c^2\over 8}+{1\over 2}
\log (\cosh (2h_1))\qquad \qquad
\hfill \hbox{for}\quad \beta <{\beta_c\over 2}
\quad \hbox{and}\quad |h_1| < {1\over 2}h_c(2\beta )
\cr
\quad &\phi_5=\beta^2 +|h_1|\tanh (h_c(2\beta ))\hfill
\quad \qquad \qquad
\hfill
\hbox{for}\quad \beta <{\beta_c\over 2} \quad
\hbox{and}\quad |h_1| > {1\over 2}h_c(2\beta )
}
$$
\medskip
\noindent
Three different ranges of temperature can be identified:
\par
\medskip
\noindent
a) If $\beta >\beta_c$ then only $\phi_3$ exists so
${\log \mid {\cal Z}\mid \over N} = \phi_3 \quad ,\forall   h$.
\par
\medskip
\noindent
b) If $\beta_c > \beta  > \beta_c/2$ we get contributions from
$\phi_1, \phi_2$  and $\phi_3$.  In this case two different regions
are found in the complex $h$  plane (figure 1, upper portion), according
to  what  term  is  the  most important.
These regions are separated by an arc which  goes  from
$(h_c(\beta ),0)$ on the real axis to $(0,h^*_2(\beta ))$ on the
imaginary axis.  In the inside of this arc $\phi_1$ dominates while in
the outside it is $\phi_3$ which is the most important. $\phi_2$ is
never relevant.
\par
\medskip
\noindent
c)  If $\beta <\beta_c/2$ then also $\phi_4$ and $\phi_5$ contribute,
and they have  to be compared to $\phi_1$ and $\phi_3$ in their
respective zones.  It  is  found that $\phi_5$ is never relevant,
i.e.    $(\phi_1-\phi_5)$ and $(\phi_3-\phi_5)$ are  never negative.
On  the  other  hand $\phi_4$  is  greater  than $\phi_3 $  (for
$h_1<{1\over 2}h_c(2\beta )$ ) and it is also greater than $\phi_1$
in  the  upper portion of the zone where $\phi_1>\phi_3$, so the line
where $\phi_4$ equals $\phi_1$ is lower than that where $\phi_3$ equals
$\phi_1$.
\par
The situation in case  c) is  that  depicted in  the  lower portion
of figure 1. The arc separating $\phi_1$ and $\phi_4$  now touches the
imaginary axis at a point $(0,h^{\hbox{\dag }}_2(\beta ))$ which
results from equating $\phi_4$ to $\phi_1$ for $h_1=0$ and satisfies
\par
$$
\cos (h^{\hbox{\dag }}_2(\beta ))=e^{-({\beta_c^2 - 2\beta^2\over 8})}
\eqno(43)
$$
\noindent
so we can see that $h^{\hbox{\dag }}_2(\beta =0)=\pi /4$.
\par
\medskip
\noindent 4.   {\bf DENSITY OF COMPLEX FIELD ZEROS OF REM}
\par
\medskip
\noindent
Having obtained $\phi ={\log \mid {\cal Z}\mid \over N}$
in the various phases of  the  model, we can calculate the density
of zeros $\rho (h)$ of ${\cal Z}$ by means  of  the formula~[13]
\par \medskip
$$
\rho (h)={1\over 2 \pi }\nabla^2\phi ={1\over 2\pi }
\left( {\partial^2\over \partial h_1^2}
+{\partial^2\over \partial h_2^2}\right) \phi (h_1,h_2)    \eqno(44)
$$
\medskip
\noindent
which is a consequence  of  the  electrostatic  analogy~[2]  that
identifies $\rho (h)$ with a charge density  and $\phi $  with  the
resulting electrostatic potential.
\par
The density of zeros associated to $\phi_1$ is  null (because the real
part of an  analytic  function  has  a  vanishing Laplacian).
The densities of zeros corresponding to $\phi_3$ and $\phi_4$ are
respectively
\par \medskip
$$
\eqalignno{
&\rho_3={1\over 2 \pi }\nabla^2\phi_3=a(1-\tanh^2(ah_1))
{\beta ^2\over \beta^2+2h_1^2(1-\tanh^2(ah_1))}  &        (45)
\cr \cr
&\rho_4={1\over 2 \pi }\nabla^2\phi_4=2 (1-\tanh^2(2h_1)) &    (46)
}
$$
\medskip
\noindent
where $a = \hat h/h_1$ as already defined.
\par \medskip \noindent
In addition to these distributed zeros, there exist other zeros
located right on the lines dividing the different zones, and
which density is proportional to the discontinuity in the
normal component of the ``electric field'' across that line. On a line
separating two generic phases $a$ and $b$ we then have
\par \medskip
$$
\rho^{ab}_l={1\over 2 \pi }\mid \bar{\nabla }(\phi_a-\phi_b)\mid
\eqno(47)
$$
\medskip
\noindent
The density of zeros on line 1-3  is found to be
\par
\medskip
$$
\rho^{13}_l={1\over 2\pi }\{ (\tanh (\hat h)-{\sinh (2h_1)\over
\cosh (2h_1)+\cos (2h2)})^2+({\sinh (2h_1)\over \cosh (2h_1)
+\cos (2h2)})^2 \}       \eqno(48)
$$
\medskip
\noindent
On line 1-4 the expression is the same with
the only difference that $\hat h(\hat h=ah_1)$ is replaced by
$2h_1$. Their zero-densities are equal where lines 1-3 and 1-4 meet.
\par
The line 3-4, which is the vertical at $h_1={1\over 2}h_c(2\beta )$
is not a line of zeros, i.e. the electric field is found to be continuous
there. On the other hand the density of {\it distributed} zeros jumps
as one crosses line 3-4 by a factor
$\{1+2(h_1/\beta )^2(1-\nobreak \tanh^2(2h_1))\}$.
\par
The density of zeros on the line 1-3 is found to be null at the point
where that line touches the real $h$ axis, in agreement with the
second order character of the field-driven transition of the model.
\par
\medskip
\noindent 5.   {\bf CONCLUSIONS}
\par
\noindent
We have estimated the behavior of the partition function of REM
in complex field. Three different phases appear: \par
For $T < T_c$ there is only one phase which we called phase 3
and is the continuation to complex field of the frozen phase of
the model. For $T_c < T$ the paramagnetic phase is also
present, which we call phase 1 (figure 1, upper portion). Finally
for $T>2T_c$ what we called phase 4 also appears at complex
values of $h$ (figure 1, lower portion).
There is no physical (i.e. for \it real  \rm values of $h$)
counterpart for this phase.   \par
We have also calculated the densities of zeros in the  different
phases. A particular structure has been revealed:  The  zeros
are  dense  near  the  real  axis  in  the  frozen   phase.  This
characteristic of the zeros distributions is intimately related to
the properties of spin glasses and has been  discussed  by  Parisi
[11], although no explicit calculation had  been  done  before.  A
numerical estimate of the complex field zeros of REM has been done
[17] and  shows  good  coincidence  [18]  with  the  results  here
reported.
\par
The existence of distributed zeros is in this  model  related
to disorder in a very clear way: The fluctuating factors $\rho _{m}$
and $\eta _{m}$
are responsible for the appearance, in $ {\cal Z} $, of nonanalyticities
of the sort which are necessary to obtain  a  non-zero  Laplacian and
consequently distributed zeros.
\par \medskip  \noindent {\bf ACKNOWLEDGEMENTS}
\par \medskip \noindent
We thank B.~Derrida for sending us his results prior to publication.
\par \medskip 
\noindent    {\bf APPENDIX}
\par
\medskip
\noindent
Here we describe the evaluation of $A_2$ by means of the steepest
descent method~[16]. Starting from
\par
$$
A_2=  \int^{m_c(\beta )}_{-m_c(\beta )} \hbox{d}m \ e^{N\Psi (m)}
\eqno(A1)
$$
\noindent
with
\par
\medskip
$$
\Psi (m)= mh+\log 2-{1\over 2}\left[ (1+m)\log (1+m)+(1-m)\log (1-m)
\right] + \left({\beta \over 2}\right)^2              \eqno(A2)
$$
\medskip
\noindent
we see that $A_2$ is non-zero only for $\beta <\beta_c$,
otherwise $m_c(\beta )=0$ and this  term does not contribute.
\par
Following the usual procedure for evaluating this kind of integrals in
the complex plane, we stretch  the  integration contour  to  infinity
following paths of steepest descent for the real part of $\Psi $ . The
imaginary part of the integrand will in this way be constant along the
whole path, and may be taken out from the integral. The level curves
of $\Psi $   are  topologically  similar  to  those  of $z^2$.  Two
steepest descent lines cross each other at the saddle point  which
is located at $m_0=\tanh (h)$: one of them goes down to  the  valleys
to the left and right of the saddle-point and will be  called  AA'
while the other climbs towards the mountain tops and is the line LL'.
The way in which the original integration path has to be
deformed depends on the relative positions of $\pm m_c$ and LL'.
\par  \noindent
Depending on $\beta $ and $h$, two situations are possible:
\par
\medskip
a)   When the integration limits $\pm m_c(\beta )$ lay on opposite sides
of LL', there are contributions from three  paths:  the  one  that
runs through the saddle-point (path  AA'),  and  those  ending  at
$\pm m_c(\beta )$.
\par
\medskip
b)   When both integration limits are on  the  same  side  of
LL', the path that joins the valleys through the saddle-point does
not contribute to the integral.
\par
\medskip  \noindent
The limit condition for this two cases  is  that $m_c(\beta )$  lays
exactly on LL', which in turn is equivalent to the condition  that
Imag$(\Psi (m_c)) = $Imag$(\Psi (m_0))$. This last identity results
from  the fact that LL', being a steepest descent for the {\it real}
part of $\Psi $, is a level curve for the {\it imaginary} part of
$\Psi $. This limit condition is then written
\par \medskip
$$
\tanh ( h_1) = {\tan(m_c(\beta )h_2) \over tg( h_2)}   =
\tanh (h^{\hbox{lim}}_1)        \eqno(A3)
$$
\medskip
\noindent
and for $\beta <\beta_c$ it defines a smooth arc in the  complex $h$
plane (with $h_1$ and $h_2$ positive) that  joins  the  points
$(h_c(\beta ),0)$  and $(0,\pi /2)$. The shape and size of this arc
depend on $\beta $. In its inner part situation a) is valid so for large
$N$
\par
\medskip
$$
A_2 \sim \exp N \{{\beta^2\over 2} + h m_c(\beta )  \} +
\exp  N \{ {\beta^2+ \beta_c^2 \over 4} + \log (\cosh (h)) \}
\eqno(A4)
$$
\medskip
\noindent
while in the outer side of this arc  b)  holds and
\par
\medskip
$$
A_2 \sim \exp N \{{\beta^2\over 2} + h m_c(\beta )  \}
\eqno(A5)
$$
\par \bigskip
\noindent {\bf REFERENCES}
\par \noindent
\parindent = -15 pt
[1]  C.~N.~Yang and T.~D.~Lee Phys.~Rev.~{\bf 87}(1952),404
\par
[2]  C.~N.~Yang and T.~D.~Lee Phys.~Rev.~{\bf 87}(1952),410
\par
[3]  Fisher M. in {\it Lectures in Theoretical Physics} {\bf 7C}
(University of Colorado Press, Boulder, Colorado, 1965)
\par
[4]  G.~Jones J.~Math.~Phys.~{\bf 7}(1966),2000
\par
[5]  van Saarloos W. and Kurtze D J.~Phys.~A
:Math.~Gen.~{\bf 17},1301(1984)
\par \noindent
J.~Stephenson and R.~Couzens Physica A {\bf 129} (1984),201
\par
[6]  Suzuki M.~Prog.~Theor.~Phys.~{\bf 38} (1967), 1243, 1255
\par  \noindent
Ono S.~et al J.~Phys.~Soc.~Japan {\bf 26s} (1969), 96
\par  \noindent
Abe R.~Prog.~Theor.~Phys.~{\bf 38} (1967), 72 , 322
\par
[7]  C.~Itzykson et al Nuc.~Phys.~{\bf B220}[FS8] (1983), 415
\par
[8]  M.~L.~Glasser et al.  Phys.~Rev.~B {\bf 35} (1987), 1841
\par
[9]  G.~Bhanot et al.~ Phys.~Rev.~Lett.~{\bf 59} (1987), 803
\par \noindent
E.~Marinari Nuc.~Phys.~{\bf B235} [FS11] (1984), 123
\par \noindent
N.~Alves et al.~ Phys.~Rev.~Lett.~ {\bf 64} (1990), 3107
\par
[10] {\it Spin Glasses and Beyond}. Mezard M.~, Parisi G.~and
Virasoro M.
\par \noindent
World Scientific. Lecture Notes in Phys.~Vol 9
\par
[11] G.~Parisi, in {\it Disordered  systems  and  Localization}
\par \noindent
Springer Lecture Notes in Phys.~Vol.~{\bf 149}.
\par
[12] B.~Derrida Phys.~Rev.~B {\bf 24} (1981), 2613
\par \noindent
B.~Derrida, Phys.~Rep.~{\bf 67} (1980), 29.
\par
[13] B.~Derrida (1991) Physica A {\bf 177} (1991), 31.
\par
[14] C.Moukarzel and N.Parga, Proceedings of the  {\it Second
  Latin American Workshop \par on Nonlinear Phenomena }, Santiago 
de Chile, September 1990 (North Holland 1991).
\par
[15] C.Moukarzel and N.Parga , Physica A {\bf 177} (1991), 24
\par
[16] {\it Asymptotic Expansions} E.~T.~Copson (Cambridge Univ.~Press,
1971)
\par
[17] C.Moukarzel and N.Parga (1991), unpublished.
\par
[18] C.Moukarzel and N.Parga (1991) {\it The REM Zeros in  the  Complex
Temperature  \par and  Magnetic  Field  Planes.}, Physica A {\bf 185} (1992), 305.
\par \bigskip
\parindent = 20 pt
\noindent {\bf FIGURE CAPTIONS}
\par
\medskip
\noindent
FIGURE 1. Shown are the  phases of the model in
the complex field plane for two different values of the temperature. For
$T_{c}<T<2T_{c}$(upper  figure) two distinct zones appear separated
by an arc of zeros. Above  this arc (phase 3) densely distributed
zeros are found while the region below it (phase 1)
is free of zeros. For $T>2T_{c}$  the outer
part of the arc appears divided in  two zones by a vertical  line,
which is not  a  line  of  zeros.  The   densities  of  zeros  are
different to the left (phase 4)
and right (phase 3) of this vertical  line.
\par
\end